\input epsf

\documentstyle[12pt]{article}

\newcommand{\sect}[1]{\setcounter{equation}{0}\section{#1}}

\def\RN{Reissner-Nordstr\"om}
\def\KK{Kaluza-Klein}
\def\am{asymptotically Minkowskian}
\def\GPS{Gross-Perry-Sorkin}

\begin{document}

\title{THE ROTATING DYONIC BLACK HOLES OF KALUZA-KLEIN THEORY}

\author{Dean Rasheed*\\DAMTP\\Silver Street\\Cambridge\\CB3 9EW}

\maketitle

\hfil{\bf Abstract\,\,\,\,\,\,}\hfil

\vbox{\tenrm\narrower
\bigskip
\noindent
The most general electrically and magnetically charged rotating black hole
solutions
of 5 dimensional \KK\ theory are given in an explicit form. Various classical
quantities associated with the black holes are derived. In particular, one
finds the
very surprising result that the gyromagnetic and gyroelectric ratios can become
{\tenit
arbitrarily large}. The thermodynamic quantities of the black holes are
calculated and
a Smarr-type formula is obtained leading to a generalized first law of black
hole
thermodynamics. The properties of the extreme solutions are investigated and it
is
shown how they naturally separate into two classes. The extreme solutions in
one class
are found to have two unusual properties: (i). Their event horizons have zero
angular
velocity and yet they have non-zero ADM angular momentum. (ii). In certain
circumstances it is possible to add angular momentum to these extreme solutions
without changing the mass or charges and yet still maintain an extreme
solution.
Regarding the extreme black holes as elementary particles, their stability is
discussed and it is found that they are stable provided they have sufficient
angular
momentum.
}

\vfil
\noindent
*dar17@amtp.cam.ac.uk

\renewcommand{\thepage}{ }
\pagebreak

\renewcommand{\thepage}{\arabic{page}}
\setcounter{page}{1}

\sect{Introduction}

In this paper we will investigate the rotating dyonic black holes of 5
dimensional
\KK\ theory. Although the original 5 dimensional theory as it stands is not a
realistic theory of nature, it continues to give insight into more
sophisticated
theories such as string theory and supergravity. The most elegant feature of
\KK\
theory is the way in which the process of dimensional reduction leads naturally
to
electromagnetism coupled to 4 dimensional gravity without the need for the
introduction of a source term on the right hand side of Einstein's equations.
All that
is needed is the assumption of an extra fifth dimension which is assumed to be
curled
up to form a circle whose radius $R_{_{KK}}$ is too small to be observed. The
existence of extra spacetime dimensions has become an integral part of many
theories
in modern theoretical physics such as string theory.

\KK\ theory arises naturally in string theory and some of the \KK\ monopoles
have been
shown to correspond to exact solutions in string theory \cite{HorTse}. The
monopoles
may also be regarded as solutions of the $N=8$ supersymmetric theory in 5
dimensions
and they fit the same supermultiplets as the original fields of the $N=8$
theory
\cite{GibPer}. \KK\ theory has also been of interest recently in connection
with
noncommutative differential geometry \cite{LanViet} which may be viewed as \KK\
theory
in which the extra fifth dimension is taken to be a discrete set of points
rather than
a continuum.

Pure gravity in 4 dimensions admits a 2 parameter family of stationary,
axi-symmetric
black hole solutions, the Kerr solutions. The 2 parameters may be chosen to be
the
mass $M$ and the angular momentum $J$ and the  condition \hbox{$M^2\ge \mid
J\mid$}
ensures cosmic censorship. When coupled to a single U(1) Maxwell field the
solutions
generalize to the Kerr-Newman family described by an additional 2 parameters,
the
electric and magnetic charges $Q$ and $P$. The condition for the singularity to
be
hidden behind an event horizon then becomes $M^2\ge Q^2+P^2+J^2/M^2$. These are
the
most general axi-symmetric black hole solutions of Einstein-Maxwell theory.
This
theory, however, requires the addition of a source term on the right hand side
of the
Einstein equations. The theory developed by Kaluza \cite{Kal} and Klein
\cite{Kle}
provides a way of unifying 4 dimensional gravity with electromagnetism without
the
need for such a source term. In this theory spacetime is considered as 5
dimensional
and dimensional reduction of the 5 dimensional vacuum Einstein equations then
leads to
4 dimensional gravity coupled to a U(1) Maxwell field and a scalar dilaton
field.

The strength of the coupling of the dilaton is fixed by the dimensional
reduction
process. It is also of interest, particularly in string theory, to consider
more
general dilaton couplings. The simplest extension of Einstein-Maxwell theory
coupled
to a scalar dilaton field $\sigma$ with coupling constant $b$ is described by
the
action
\begin{equation}
S = \int d^4x \sqrt{g} \left[ R - 2(\partial\sigma)^2 - e^{2b\sigma}F^2 \right]
\label{DilAction}
\end{equation}
which leads to the following equations of motion
\begin{eqnarray}
& R_{\mu\nu} = 2(\partial\sigma_\mu)(\partial\sigma_\nu) +
2e^{2b\sigma}T_{\mu\nu}
\nonumber \\
& \nonumber \\
& \nabla_\mu(e^{2b\sigma}F^{\mu\nu}) = 0 \label{EoM} \\
& \nonumber \\
& \Box\sigma = {b\over 2}e^{2b\sigma}F^2 \nonumber
\end{eqnarray}
where $T_{\mu\nu}$ is the energy-momentum tensor of the Maxwell field
\begin{equation}
T_{\mu\nu} = F_{\mu\rho}{F_\nu}^\rho - {1\over 4}g_{\mu\nu}F^2.
\end{equation}
When $b=0$ this reduces to Einstein-Maxwell theory. For $b\neq 0$ it is
possible to
consistently set $\sigma\equiv 0$ in (\ref{EoM}) only when $F^2=0$. This is the
Einstein-Maxwell embedding and includes the $Q=P$ \RN\ solutions but not, in
general,
the $Q=P$ Kerr-Newman solutions.

The $b=1$ case of (\ref{DilAction}) arises naturally in string theory and the
$b=\sqrt{3}$ case is the one given by \KK\ theory which will be the main
subject of
this paper. Less is known about other values of the dilaton coupling. The black
hole
solutions of this theory will have an additional charge $\Sigma$, the scalar
dilaton
charge given by
\begin{equation}
\sigma \sim {\Sigma\over r} \qquad {\rm as} \quad r\rightarrow\infty.
\end{equation}
In general this leads to the event horizon becoming singular unless $\Sigma$
takes a
specific value determined by the other charges. Thus the stationary black hole
solutions can still be labelled by the 4 parameters $M$,$P$,$Q$ and $J$. The
electrically charged static solutions for general $b$ are known \cite{GibMad},
\cite{GarHor}. These have been generalized to slowly rotating solutions by
expanding
(\ref{EoM}) linearly in angular momentum \cite{HorHor}.

The aim of this paper is to find the most general axi-symmetric black hole
solutions
of the $b=\sqrt{3}$ \KK\ theory so that they may be compared and contrasted
with those
of Einstein-Maxwell theory and string theory.

\sect{\KK\ theory}

The vacuum Einstein equations in 5 dimensions can be derived from the action
\begin{equation}
S = \int d^5x \sqrt{^{(5)}g}\,\, {^{(5)}R}.
\label{5DAction}
\end{equation}
The extra coordinate $x^5$ is assumed to be periodic with period $2\pi
R_{_{KK}}$ and
in addition ${\partial\over\partial x^5}$ is assumed to be killing so that the
5
dimensional metric components are functions of $x^\mu\,(\mu=0\dots 3)$ only.
The 5
dimensional metric can be written in the form
\begin{equation}
ds_{(5)}^2 = e^{4\sigma/\sqrt{3}}\left(dx^5+2A_\mu dx^\mu\right)^2 +
e^{-2\sigma/\sqrt{3}}g_{\mu\nu}dx^\mu dx^\nu
\label{5DMetric}
\end{equation}
and the action (\ref{5DAction}) then reduces to
\begin{equation}
S = 2\pi R_{_{KK}}\int d^4x \sqrt{g} \left[R - 2(\partial\sigma)^2 -
e^{2\sigma\sqrt{3}}F^2\right]
\end{equation}
where $F_{\mu\nu}=\partial_\mu A_\nu-\partial_\nu A_\mu$. This is equivalent to
the
$b=\sqrt{3}$ case of (\ref{DilAction}).

When $\xi={\partial\over\partial t}$ is also killing the 5 dimensional metric
can be
further decomposed as
\begin{equation}
ds_{(5)}^2 =
\lambda_{ab}\left(dx^a+{\omega^a}_idx^i\right)\left(dx^b+{\omega^b}_jdx^j\right) +
{1\over\tau}h_{ij}dx^idx^j
\label{Metric}
\end{equation}
where $\tau=-{\rm det}\,\lambda_{ab}$. $a,b,\dots$ take the values 0 and 5, and
$i,j,\dots$ run from 1 to 3. $\lambda_{ab}$, ${\omega^a}_i$ and $h_{ij}$ are
functions
of the 3 spatial coordinates $x^i$. The $R_{ai}$ components of the vacuum
Einstein
equations imply that the ${\omega^a}_i$ can be expressed in terms of {\em twist
potentials} $V_a$ satisfying
\begin{equation}
V_{a,i} = \tau\lambda_{ab}{\varepsilon_i}^{jk}{\omega^b}_{j,k}
\label{Twists}
\end{equation}
where $\varepsilon_{ijk}$ is the completely antisymmetric {\em tensor} of the 3
dimensional metric $h_{ij}$. Then, defining the symmetric, unimodular matrix
\begin{equation}
\chi = \left(
\begin{array}{cc}
\lambda_{ab}-{1\over\tau}V_aV_b & {1\over\tau}V_a \\
 & \\
{1\over\tau}V_b & -{1\over\tau}
\end{array}
\right)
\label{Chi}
\end{equation}
the remaining Einstein equations can be written as \cite{Maison}
\begin{eqnarray}
& \left(\chi^{-1}\chi^{,i}\right)_{;i} = 0 \nonumber \\
& \label{ChiEqns} \\
& R_{ij} = {1\over 4}{\rm
Tr}\,\left(\chi^{-1}\chi_{,i}\chi^{-1}\chi_{,j}\right)
\nonumber
\end{eqnarray}
where $;$ denotes the covariant derivative with respect to $h_{ij}$. These
equations
can be derived from the 3 dimensional $\sigma$-model action
\begin{equation}
S = \int d^3x \sqrt{h} \left[{^{(3)}}R - {1\over 4}{\rm
Tr}\,\left(\chi^{-1}\chi_{,i}\chi^{-1}\chi^{,i}\right)\right].
\end{equation}
Clearly equations (\ref{ChiEqns}) are invariant under SL(3,$I\!\!R$)
transformations.
Since $\chi$ is a symmetric matrix the most natural group action to consider is
\begin{equation}
\chi \mapsto N\chi N^T \quad N\in{\rm SL(3,}I\!\!R{\rm )}.
\end{equation}
In order for $\chi$ to represent an \am\ spacetime it is necessary that
\begin{equation}
\chi \rightarrow \eta = \left(
\begin{array}{ccc}
-1 & & \\
 & 1 & \\
 & & -1
\end{array}
\right) \quad {\rm as} \quad r \rightarrow \infty
\label{AsymMink}
\end{equation}
and only the subgroup SO(1,2) of SL(3,$I\!\!R$) transformations preserves this
property, i.e.\ those $N$ satisfying $N^{-1}=\eta N^T\eta$. It is thus possible
to
transform continuously through the space of solutions obtaining new black hole
solutions from old ones by applying SO(1,2) transformations to $\chi$.

The only complication is that matrices $\chi$ of the form (\ref{AsymMink}) may
not
represent \am\ spacetimes (in either 4 or 5 dimensions). Indeed the
magnetically
charged solutions we seek will not be \am\ from the 5 dimensional point of
view,
however some SO(1,2) transformations will lead to solutions that are not even
\am\ in
4 dimensions. These solutions are Taub-NUT like solutions in 4 dimensions. In
order to
generate 4 dimensional black hole solutions it will be necessary to further
restrict
the transformations that can be applied to $\chi$. The restricted
transformations,
which no longer form a group, will be labelled by the 2 parameters
$\alpha,\beta\in
I\!\!R$.

\sect{Static Solutions}

The simplest solutions of (\ref{ChiEqns}) take the form
\begin{equation}
\chi = \eta e^{Af(x^i)}
\label{Geodesic}
\end{equation}
where $A$ is a constant matrix and $f$ is required to be a harmonic function of
the
spatial coordinates $x^i$ with respect to the 3 dimensional metric $h_{ij}$.
These are
geodesics in the symmetric space SL(3,$I\!\!R$)/SO(3) with metric $dS^2={\rm
Tr}\,\left(\chi^{-1}d\chi\,\chi^{-1}d\chi\right)$ \cite{Clem}. The requirement
that
$\chi$ be symmetric and unimodular imposes 2 constraints on the matrix $A$:
\begin{eqnarray}
& {\rm Tr}\,A = 0 \nonumber \\
& \label{AConstraint} \\
& A^T =  \eta A\eta. \nonumber
\end{eqnarray}
The most general matrix satisfying these constraints may be written in the
suggestive
form
\begin{equation}
A = \left(
\begin{array}{ccc}
-2M-2\Sigma/\sqrt{3} & -2Q & 2N \\
 & & \\
2Q & 4\Sigma/\sqrt{3} & 2P \\
 & & \\
2N & -2P & 2M-2\Sigma/\sqrt{3}
\end{array}
\right).
\label{GenA}
\end{equation}
If $f\sim{1\over r}$, $N$ will be related to the NUT charge of the 4
dimensional
spacetime. We will take $N$ to be zero from now on. In that case $M$ is the ADM
mass
of the 4 dimensional spacetime and $P$, $Q$ and $\Sigma$ are the magnetic,
electric
and scalar charges respectively.

Null geodesics in the symmetric space are given by the further constraint
\begin{equation}
{\rm Tr}\,\left(A^2\right) = 0
\end{equation}
which is equivalent to Scherk's antigravity condition \cite{Scherk},
\cite{Gibb}
\begin{equation}
M^2 + \Sigma^2 = P^2 + Q^2.
\label{Null}
\end{equation}
which ensures a force balance between monopoles allowing for the possible
existence of
static multi-centre solutions. These are extreme black holes whose 3
dimensional
metric $h_{ij}$ is Ricci flat but which are only flat in the 3 cases described
below
-- the extreme electric (plane wave) solution, the \GPS\ magnetic monopole and
the
extreme $P=Q$ \RN\ embedding.

The SO(1,2) transformation of $\chi$
\begin{equation}
\chi \rightarrow N\chi N^T \quad,\quad N^{-1} = \eta N^T\eta
\label{ChiTransf}
\end{equation}
corresponds to a similarity transformation of the matrix A
\begin{equation}
A \rightarrow MAM^{-1} \quad,\quad M = \eta N\eta\,\in{\rm SO(1,2).}
\label{Atransf}
\end{equation}
There are thus 2 natural classes of solutions of the form (\ref{Geodesic}) to
be
considered, depending on whether $A$ is singular or non-singular. Consider
first the
case of non-singular $A$. All matrices in this class are similar to the
traceless
diagonal matrix
\begin{equation}
A = \left(
\begin{array}{ccc}
-2M-2\Sigma/\sqrt{3} & & \\
 & & \\
 & 4\Sigma/\sqrt{3} & \\
 & & \\
 & & 2M-2\Sigma/\sqrt{3}
\end{array}
\right).
\end{equation}
Exponentiating this matrix and solving for the metric components using
(\ref{Metric}),
(\ref{Twists}), (\ref{Chi}), (\ref{ChiEqns}) and (\ref{Geodesic}) gives the 5
dimensional metric. The 4 dimensional metric, extracted using (\ref{5DMetric}),
is
\begin{eqnarray}
&ds_{(4)}^2 = -\left(1-{2{\tilde M}\over r}\right)^{M/{\tilde M}}dt^2 +
\left(1-{2{\tilde M}\over r}\right)^{-M/{\tilde M}}\left[dr^2 +
r^2\left(1-{2{\tilde
M}\over r}\right)d\Omega^2\right], \nonumber \\
 \\
&{\tilde M} = \sqrt{M^2+\Sigma^2}. \nonumber
\end{eqnarray}
When $\Sigma=0$ this is just the Schwarzschild solution and for $\Sigma\neq 0$
it is
asymptotically like the Schwarzschild solution. However, when $\Sigma\neq 0$,
the
horizon at $r=2\sqrt{M^2+\Sigma^2}$ becomes singular and it is expected that
all
solutions in this class with ${\rm det}\,A\neq 0$ will be singular in this way.

Now consider those matrices of the form (\ref{GenA}) which are singular. The
condition
${\rm det}\,A=0$ gives a cubic equation for $\Sigma$ in terms of the mass $M$
and the
other charges $P$ and $Q$:
\begin{equation}
{Q^2\over\Sigma+M\sqrt{3}} + {P^2\over\Sigma-M\sqrt{3}} = {2\Sigma\over 3}.
\label{Cubic}
\end{equation}
The general spherically symmetric solution in this class depending on one
harmonic
function $f$ may be obtained from the Schwarzschild solution by applying a
restricted
2 parameter family of SO(1,2) transformations that do not introduce a NUT
charge. The
Schwarzschild solution is represented by the matrix
\begin{equation}
A = \left(
\begin{array}{ccc}
-2M & & \\
 & 0 & \\
 & & 2M
\end{array}
\right).
\end{equation}
The solutions obtained from this correspond to the static, spherically
symmetric
dyonic solutions obtained in \cite{Leut}--\cite{Poll} and which were thoroughly
investigated in \cite{GibWil}. These solutions are a special case of the more
general
rotating solutions which will be given in the next section and so we will
postpone any
detailed discussion of them and their derivation until then. A few important
special
cases are worth mentioning:

\bigskip
\noindent
(1). The 5 dimensional plane wave solution
\begin{equation}
ds_{(5)}^2 = \left(1+{4M\over r}\right)\left(dx^5\right)^2 + 2dx^5dt +
\mbox{\boldmath
$dx.dx$}
\label{PPwave}
\end{equation}
which represents the extreme electric \KK\ monopole with $Q=2M$,$P=0$ and
$\Sigma=M\sqrt{3}$ has
\begin{equation}
A = \left(
\begin{array}{ccc}
-4M & -4M & 0 \\
4M & 4M & 0 \\
0 & 0 & 0
\end{array}
\right) \quad,\quad h_{ij}=\delta_{ij} \quad,\quad f={1\over r}.
\end{equation}

\bigskip
\noindent
(2). The \GPS\ magnetic monopole \cite{GroPer}, \cite{Sork}
\begin{equation}
ds_{(5)}^2 = -dt^2 + {1\over 1+{4M\over r}}\left(dx^5-4M\cos\theta
d\phi\right)^2 +
\left(1+{4M\over r}\right)\mbox{\boldmath $dx.dx$}
\label{gps}
\end{equation}
with $P=2M$, $Q=0$, $\Sigma=-M\sqrt{3}$,
\begin{equation}
A = \left(
\begin{array}{ccc}
0 & 0 & 0 \\
0 & -4M & 4M \\
0 & -4M & 4M
\end{array}
\right) \quad,\quad h_{ij}=\delta_{ij} \quad{\rm and}\quad f={1\over r}.
\end{equation}
This is the electromagnetic dual of the previous solution where the discrete
electromagnetic duality transformation
\begin{equation}
e^{2\sigma\sqrt{3}}F_{\mu\nu} \rightarrow *F_{\mu\nu} \quad,\quad \sigma
\rightarrow
-\sigma \quad,\quad g_{\mu\nu} \rightarrow g_{\mu\nu}
\label{e-m}
\end{equation}
is a symmetry of (\ref{EoM}) and, in the non-rotating case, is equivalent to
exchanging $P$ and $Q$, and changing the sign of $\Sigma$.

\bigskip
\noindent
(3). The extreme \RN\ embedding
\begin{eqnarray}
& ds_{(5)}^2 = \left(dx^5-M\sqrt{2}\cos\theta d\phi+{M\sqrt{2}\over
r}dt\right)^2 -
\left(1-{m\over r}\right)^2dt^2 \nonumber \\
\label{RNemb} \\
& + \left(1-{m\over r}\right)^{-2}dr^2 + r^2d\Omega^2 \nonumber
\end{eqnarray}
with $P=Q=M/\sqrt{2}$, $\Sigma=0$,
\begin{equation}
A = \left(
\begin{array}{ccc}
-2M & -M\sqrt{2} & 0 \\
M\sqrt{2} & 0 & M\sqrt{2} \\
0 & -M\sqrt{2} & 2M
\end{array}
\right) \quad,\quad h_{ij}=\delta_{ij} \quad{\rm and}\quad f={1\over r}.
\end{equation}

\bigskip
\bigskip
\noindent
These 3 solutions all have a flat 3 dimensional metric $h_{ij}$ and so are easy
to
generalize to multi-centre solutions, simply by replacing $f$ by
\begin{equation}
f = \sum_{i=1}^n {\lambda_i\over\mid\mbox{\boldmath $x$}-\mbox{\boldmath
$x$}_i\mid}
\end{equation}
which gives $n$ monopoles of masses $\lambda_iM$ at $\mbox{\boldmath
$x$}=\mbox{\boldmath $x$}_i$. By explicitly solving equations (\ref{Twists})
and
(\ref{ChiEqns}) it can be shown that these 3 solutions are the only spherically
symmetric solutions with flat spatial metric $h_{ij}$ and so these will be the
only
multi-centre solutions of this form.

An important special case of an SO(1,2) transformation that can be applied to
the
matrix $A$ is
\begin{equation}
M = \left(
\begin{array}{ccc}
\cosh\alpha & \sinh\alpha & 0 \\
\sinh\alpha & \cosh\alpha & 0 \\
0 & 0 & 1
\end{array}
\right)
\end{equation}
which corresponds to applying the Lorentz boost
\begin{equation}
\left\{
\begin{array}{rl}
x^5 &\rightarrow\gamma\left(x^5+vt\right) \\
 & \\
t &\rightarrow\gamma\left(t+vx^5\right)
\end{array}
\right.\qquad{\rm where}\quad \gamma=\left(1-v^2\right)^{-{1\over
2}}=\cosh\alpha.
\label{LBoost}
\end{equation}
This gives a particularly easy way of obtaining the purely electrically charged
solutions from neutral ones and it generalizes easily to the rotating case
\cite{GibWil}, \cite{HorHor} and \cite{FroZel}. The extreme electric \KK\
solution
(\ref{PPwave}) above corresponds to the limit $v\rightarrow 1$ and thus
represents a
Schwarzschild black hole moving at the speed of light in 5 dimensions.

The extreme solutions satisfy both (\ref{Null}) and (\ref{Cubic}) which, on
eliminating $\Sigma$ is equivalent to the astroid
\begin{equation}
\left({Q\over M}\right)^{2\over 3} + \left({P\over M}\right)^{2\over 3} =
2^{2\over 3}.
\label{Ast}
\end{equation}
This is to be compared with the curve of spherically symmetric extreme
solutions in
Einstein-Maxwell theory ($b=0$)
\begin{equation}
\left({Q\over M}\right)^2 + \left({P\over M}\right)^2 = 1
\end{equation}
and in string theory ($b=1$)
\begin{equation}
\left|{Q\over M}\right| + \left|{P\over M}\right| = \sqrt{2}.
\end{equation}
So these different values of dilaton coupling $b$ fit a family of power law
curves of
the form
\begin{equation}
\left|{Q\over M}\right|^n + \left|{P\over M}\right|^n = K^n.
\end{equation}
In the pure electric case, the extreme solution saturates the Bogomol'nyi bound
\cite{GibKas}
\begin{equation}
\left|{Q\over M}\right| = \sqrt{1+b^2}
\end{equation}
and so $K=\sqrt{1+b^2}$. The $P=Q=M/\sqrt{2}$ \RN\ solution is an extreme
solution for
all values of the dilaton coupling $b$ and this fixes $n$ as a function of $b$.
So the
curves of extreme solutions for different values of b can be summarized as
\begin{equation}
\left\{
\begin{array}{l}
\bigl|{Q\over M}\bigr|^n + \bigl|{P\over M}\bigr|^n =
\left(1+b^2\right)^{n\over 2} \\
\\
n = {2\over 1+\log_2\left(1+b^2\right)}
\end{array}
\right.
\label{GenDil}
\end{equation}
which is a family of power law curves touching at one point, see Fig.~1. The
formulae
(\ref{GenDil}) are certainly true for $b=0,1,\sqrt{3}$ and they may also be
true more
generally.

\begin{figure}
\epsffile{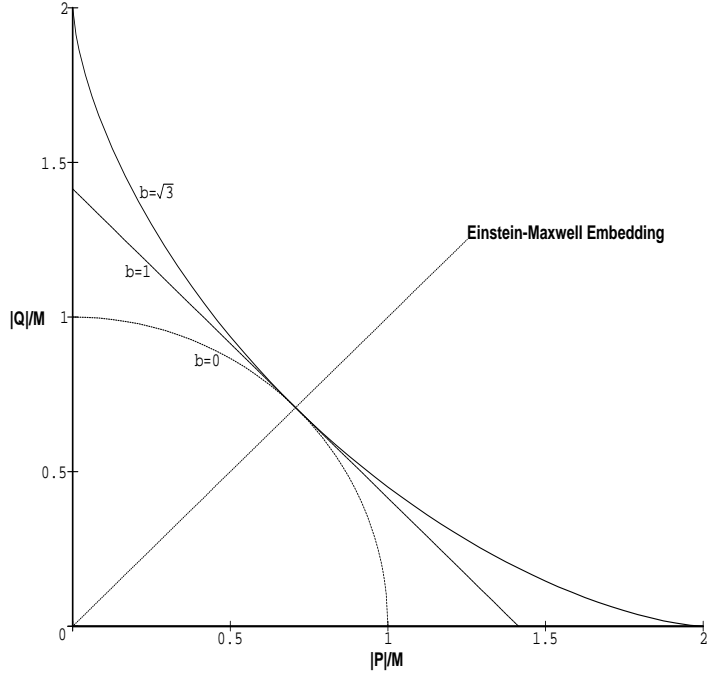}
\caption{Extreme Solutions for Different Dilaton Couplings $b$}
\end{figure}

\sect{Rotating Solutions}

A class of rotating solutions may be represented by totally geodesic surfaces
in the
symmetric space depending on 2 harmonic functions $f(x^i)$ and $g(x^i)$
\cite{Clem}
\begin{equation}
\chi = \eta e^{Af}e^{Bg}
\end{equation}
where $A$ and $B$ are constant, commuting matrices, both satisfying
(\ref{AConstraint}). This turns out not to be a convenient representation of
the
solutions, however. It is possible to generate all the rotating solutions from
the
Kerr solution by acting on the matrix $\chi$ corresponding to the Kerr solution
with a
restricted set of SO(1,2) transformations of the form (\ref{ChiTransf}). The
general
rotating solution is expected to have a $\chi$-matrix which is asymptotically
of the
form
\begin{equation}
\chi \sim \left(
\begin{array}{ccc}
-1+{2M+2\Sigma/\sqrt{3}\over r} & {2Q\over r} & -{2J\cos\theta\over r^2} \\
 & & \\
{2Q\over r} & 1+ {4\Sigma\over r\sqrt{3}} & {2P\over r} \\
 & & \\
-{2J\cos\theta\over r^2} & {2P\over r} & -1-{2M-2\Sigma/\sqrt{3}\over r}
\end{array}
\right).
\label{ChiAsym}
\end{equation}
Note that the (1,3) component is required to be O(${1\over r^2}$) in order that
the 4
dimensional metric has no NUT charge and is asymptotically flat. This condition
is
{\em not} preserved by general SO(1,2) transformations (\ref{ChiTransf}) thus
close
attention must be payed to precisely which transformations {\em can} be applied
to the
Kerr solution to obtain other black hole solutions.

The Kerr solution of mass $M_K$ and angular momentum $J_K=aM_K$ is
$$ds_{(5)}^2 = \left(dx^5\right)^2 - \left(1-Z\right)\left(dt +
{aZ\sin^2\theta\over
1-Z} d\phi\right)^2 + {\rho\over\Delta}dr^2 + \rho d\theta^2 + {\Delta\over
1-Z}\sin^2\theta d\phi^2$$
where
\begin{equation}
\left\{
\begin{array}{l}
\rho = r^2 + a^2\cos^2\theta \\
\\
\Delta = r^2 - 2M_Kr + a^2 \\
\\
Z = {2M_Kr\over\rho}.
\end{array}
\right.
\label{Kerr}
\end{equation}
\pagebreak[1]
Therefore
\begin{eqnarray}
\lambda_{ab} = \left(
\begin{array}{cc}
-(1-Z) & 0 \\
0 & 1
\end{array}
\right) \quad,\quad \tau = 1-Z \nonumber \\
\\
\mbox{\boldmath $\omega$}^0\mbox{\boldmath $.dx$} = {aZ\sin^2\theta\over 1-Z}
d\phi
\quad,\quad \mbox{\boldmath $\omega$}^5 = 0. \nonumber
\end{eqnarray}
The twist potentials are then
\begin{equation}
V_0 = {-2M_Ka\cos\theta\over\rho} \quad,\quad V_5 = 0
\end{equation}
so the $\chi$-matrix for the Kerr solution is
\begin{equation}
\chi_{_K} = \left(
\begin{array}{ccc}
-(1-Z)-{4M_K^2a^2\cos^2\theta\over\rho^2(1-Z)} & 0 & -{2M_Ka\cos\theta\over\rho
(1-Z)}
\\
 & & \\
0 & 1 & 0 \\
 & & \\
-{2M_Ka\cos\theta\over\rho (1-Z)} & 0 & -{1\over 1-Z}
\end{array}
\right).
\label{ChiKerr}
\end{equation}
We now apply an SO(1,2) transformation to this:
\begin{equation}
\chi = N\chi_{_K}N^T
\label{NewChi}
\end{equation}
but restricted to those transformations $N$ that preserve the asymptotic form
(\ref{ChiAsym}). First we decompose the general SO(1,2) transformation into 2
boosts
and a rotation, $N=N_1N_2N_3$, where
\begin{equation}
\left\{
\begin{array}{l}
N_1 = \left(
\begin{array}{ccc}
\cosh\alpha & \sinh\alpha & 0 \\
\sinh\alpha & \cosh\alpha & 0 \\
0 & 0 & 1
\end{array}
\right), \\
\\
N_2 = \left(
\begin{array}{ccc}
1 & 0 & 0 \\
0 & \cosh\beta & \sinh\beta \\
0 & \sinh\beta & \sinh\beta
\end{array}
\right), \\
\\
N_3 = \left(
\begin{array}{ccc}
\cos\gamma & 0 & -\sin\gamma \\
0 & 1 & 0 \\
\sin\gamma & 0 & \cos\gamma
\end{array}
\right).
\end{array}
\right.
\label{Boost}
\end{equation}
The requirement that $\chi$ have the asymptotic form (\ref{ChiAsym}) gives the
following constraint on the matrix $N$
\begin{equation}
N_{(1,1)}N_{(3,1)} = N_{(1,3)}N_{(3,3)}
\label{NConstraint}
\end{equation}
which is equivalent to
\begin{equation}
\tan 2\gamma = \tanh\alpha\sinh\beta.
\label{GammaConstraint}
\end{equation}
Thus the allowed SO(1,2) transformations can be parametrized by the 2 boost
parameters
$\alpha$,$\beta\in I\!\!R$. Note that these restrictions (\ref{NConstraint}),
(\ref{GammaConstraint}) giving the allowed transformation matrices $N$ are
specific to
the Kerr solution. A matrix $N$ applied to a different solution would have a
different
restriction on it. This is equivalent to the statement that the restricted set
of
transformations that do not introduce a NUT charge in the 4 dimensional metric
do not
form a group.

The transformed matrix $\chi$ can now be calculated in terms of $\alpha$,
$\beta$ and
$\chi_{_K}$, and the 5 dimensional metric can be reconstructed from this. Some
lengthy
algebra gives the result
\begin{equation}
ds_{(5)}^2 = {B\over A}\left(dx^5 + 2A_\mu dx^\mu\right)^2 + \sqrt{A\over
B}ds_{(4)}^2
\end{equation}
where
\begin{equation}
ds_{(4)}^2 = -{f^2\over\sqrt{AB}}\left(dt+{\omega^0}_\phi d\phi\right)^2 +
{\sqrt{AB}\over\Delta}dr^2 + \sqrt{AB}d\theta^2 + {\Delta\sqrt{AB}\over
f^2}\sin^2\theta d\phi^2
\label{GenMetric}
\end{equation}
and
\begin{eqnarray}
A = \left(r-\Sigma /\sqrt{3}\right)^2 - {2P^2\Sigma\over\Sigma - M\sqrt{3}} +
a^2\cos^2\theta + {2JPQ\cos\theta\over\left(M+\Sigma/\sqrt{3}\right)^2-Q^2},
\nonumber
\\
\nonumber \\
B = \left(r+\Sigma /\sqrt{3}\right)^2 - {2Q^2\Sigma\over\Sigma + M\sqrt{3}} +
a^2\cos^2\theta - {2JPQ\cos\theta\over\left(M-\Sigma/\sqrt{3}\right)^2-P^2},
\label{AB} \\
\nonumber \\
{\omega^0}_\phi = {2J\sin^2\theta\over
f^2}\left[r-M+{\left(M^2+\Sigma^2-P^2-Q^2\right)\left(M+\Sigma
/\sqrt{3}\right)\over\left(M+\Sigma /\sqrt{3}\right)^2-Q^2}\right]. \nonumber
\end{eqnarray}
Here the radial coordinate has been translated
\begin{equation}
r \mapsto r + M_K - M
\end{equation}
so that now
\begin{equation}
\Delta = r^2 - 2Mr + P^2 + Q^2 - \Sigma^2 + a^2
\label{Delta}
\end{equation}
and
\begin{equation}
f^2 = r^2 - 2Mr + P^2 + Q^2 - \Sigma^2 + a^2\cos^2\theta.
\label{f2}
\end{equation}
The electromagnetic vector potential is given by
\begin{equation}
2A_\mu dx^\mu = {C\over B}dt + \left({\omega^5}_\phi + {C\over
B}{\omega^0}_\phi\right)d\phi
\label{Aem}
\end{equation}
where
$$C = 2Q\left(r-\Sigma /\sqrt{3}\right) - {2PJ\cos\theta\left(M+\Sigma
/\sqrt{3}\right)\over\left(M-\Sigma /\sqrt{3}\right)^2-P^2},$$
\begin{equation}\label{C}\end{equation}
$${\omega^5}_\phi = {2P\Delta\over f^2}\cos\theta - {2QJ\sin^2\theta
\left[r\left(M -
\Sigma/\sqrt{3}\right) + M\Sigma/\sqrt{3} + \Sigma^2-P^2-Q^2\right] \over
f^2\left[\left(M+\Sigma/\sqrt{3}\right)^2-Q^2\right]}.$$
The new mass $M$, electric charge $Q$, magnetic charge $P$, new angular
momentum $J$
and dilaton charge $\Sigma$ are related to the old Kerr parameters $M_K$, $J_K$
and
the boost parameters $\alpha$, $\beta$ by
$$M = {M_K\left(1+\cosh^2\alpha\cosh^2\beta\right)\cosh\alpha\over
2\sqrt{1+\sinh^2\alpha\cosh^2\beta}},$$
\begin{equation}
\Sigma =
{\sqrt{3}M_K\cosh\alpha\left(1-\cosh^2\beta+\sinh^2\alpha\cosh^2\beta\right)\over
2\sqrt{1+\sinh^2\alpha\cosh^2\beta}},
\label{NewParam}
\end{equation}
$$Q = M_K\sinh\alpha\sqrt{1+\sinh^2\alpha\cosh^2\beta} \quad,\quad P =
{M_K\sinh\beta\cosh\beta\over\sqrt{1+\sinh^2\alpha\cosh^2\beta}},$$
$$J = aM_K\cosh\beta\sqrt{1+\sinh^2\alpha\cosh^2\beta}.$$
It is easy to check that $\Sigma$ does indeed satisfy the cubic equation
\begin{equation}
{Q^2\over\Sigma+M\sqrt{3}} + {P^2\over\Sigma-M\sqrt{3}} = {2\Sigma\over 3}
\label{Cubic2}
\end{equation}
and the Kerr mass $M_K$ is related to the parameters of the new solution by
\begin{equation}
M_K^2 = M^2 + \Sigma^2 - P^2 - Q^2.
\end{equation}
$J$ and a are not independent parameters but are related via
\begin{equation}
J^2 = a^2{\left[\left(M+\Sigma/\sqrt{3}\right)^2-Q^2\right]
\left[\left(M-\Sigma/\sqrt{3}\right)^2-P^2\right]\over M^2+\Sigma^2-P^2-Q^2}
\label{Ja}
\end{equation}

This is the general, rotating, dyonic black hole solution. It depends on 4
parameters,
$M$, $P$, $Q$ and $J$, although for some purposes it is more convenient to use
the
parameters $M_K$, $a$, $\alpha$ and $\beta$.

\sect{Extreme Solutions}

It is interesting to study the conditions under which (\ref{GenMetric})
satisfies
cosmic censorship. It turns out that the surface of extreme solutions is no
longer a
smooth surface as it is in Einstein-Maxwell theory, where it is a sphere.
Instead it
is made up of 2 distinct smooth surfaces {\bf S} and {\bf W} which intersect at
a
curve, see Fig.~2.

The horizons of (\ref{GenMetric}) are given by the zeros of $\Delta$ and so a
necessary condition for them to be present is
\begin{equation}
M^2 \ge P^2 + Q^2 + a^2 - \Sigma^2
\end{equation}
which is precisely the condition for the original Kerr solution to have
horizons,
$M_K\ge\mid a\mid$. Thus boosting the extremely rotating Kerr solution will
give (part
of) the surface of extreme rotating dyons. This is surface {\bf S} in Fig.~2.

This is only part of the surface, however, because when $M^2+\Sigma^2=P^2+Q^2$,
$a$ is
necessarily zero but (\ref{Ja}) breaks down and the angular momentum $J$ may be
non-zero. This second set of extreme solutions forms a vertical surface ({\bf
W} in
Fig.~2.) above the astroid of non-rotating extreme solutions. To see these
solutions,
consider the $\beta\rightarrow\infty$ limit of the boosted Kerr solutions. In
this
limit
\begin{equation}
{J\over M^2} = {4a\sinh^3\alpha\over M_K\cosh^6\alpha} \quad,\quad {P\over M} =
{2\over\cosh^3\alpha} \quad,\quad {Q\over M} =
{2\sinh^3\alpha\over\cosh^3\alpha}.
\end{equation}
Both $a$ and $M_K$ vanish in this limit but their ratio remains less than 1 and
so

\begin{figure}
\epsffile{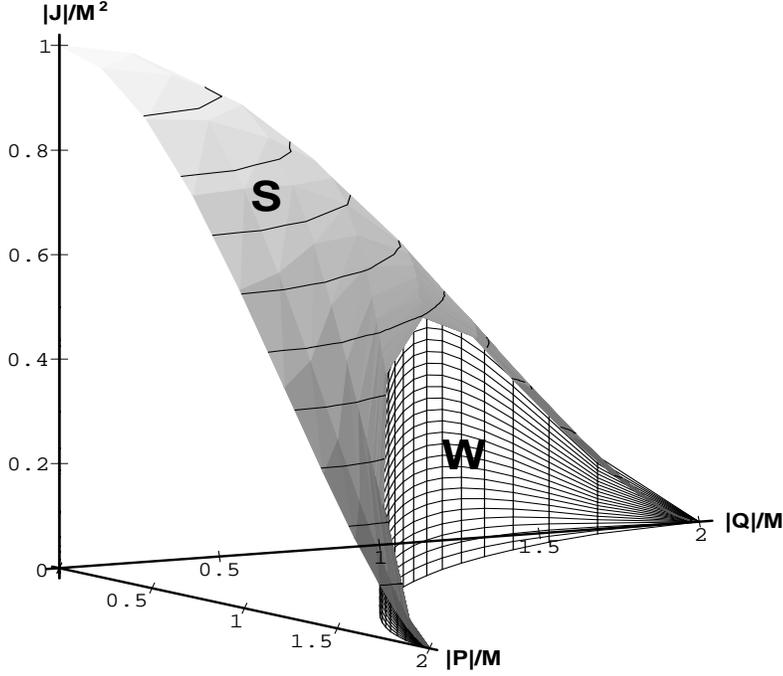}
\caption{Surfaces of Extreme Solutions in \KK\ Theory}
\end{figure}

\begin{equation}
\left({P\over M}\right)^{2\over 3} + \left({Q\over M}\right)^{2\over 3} =
2^{2\over 3} \quad,\quad J \le PQ
\label{W}
\end{equation}
which is the vertical wall {\bf W} in Fig.~2.

So we have the result that any extreme non-rotating dyon lying on the astroid
(\ref{Ast}) can be given angular momentum $J\le PQ$ with $M$,$P$ and $Q$
remaining
fixed and the solution will remain extreme. This is an unexpected result since
in
Einstein-Maxwell theory adding angular momentum to an extreme solution would
make it
ultra-extreme. This unusual property of \KK\ black holes only occurs when
dyonic black
holes are considered since the maximum amount of angular momentum that can be
added
without a naked singularity resulting is $J=PQ$. It is a subject for further
research
to discover if this is a general feature of all theories with non-zero dilaton
coupling.

To see a specific example of this effect, consider the extreme $P=Q$ solutions.
These
fall into 2 classes. First there are those solutions that which have
$P=Q=M/\sqrt{2}$
and $J\le PQ=M^2/2$. These correspond to adding angular momentum to the $P=Q$
extreme
\RN\ solution (\ref{RNemb}) and they have a particularly simple 4 dimensional
metric
\begin{eqnarray}
& ds_{(4)}^2 = -{\left(1-{M\over r}\right)^2\over
F}\left(dt-{2Ma\sin^2\theta\over
r-m}d\phi\right)^2 + {F\over\left(1-{M\over r}\right)^2}dr^2 + Fr^2d\Omega^2,
\nonumber \\
& \\
& F = \sqrt{1-{4M^2a^2\cos^2\theta\over r^4}}. \nonumber
\label{PeqQJneq0}
\end{eqnarray}
Note that for $J\ne 0$ this is no longer the Einstein-Maxwell embedding since
$F_{\mu\nu}F^{\mu\nu}$ is no longer zero. The event horizon is at $r=M$ and
there is
no ergoregion even when $J\ne 0$ (this is true of all the solutions on the
surface
{\bf W} in Fig.~2.). The singularity can be found by looking at invariants
formed from
the Riemann tensor such as $R_{\mu\nu\rho\sigma}R^{\mu\nu\rho\sigma}$ and this
tells
us that the 4 dimensional metric is everywhere non-singular except at
\begin{equation}
r = \sqrt{2\mid J\cos\theta\mid}
\end{equation}
and this singularity will be hidden safely behind the event horizon provided
that
$J\le M^2/2$. The second class of extreme $P=Q$ dyons can be obtained by
boosting the
$a=M_K$ Kerr solution as above, subject to the constraint
\begin{equation}
\sinh\alpha = \tanh\beta.
\end{equation}
The result is a curve of extreme solutions starting at the $a=M_K$ Kerr
solution
($\alpha=\beta=0$) and extending to the $P=Q=M/\sqrt{2}$, $J=M^2/2$ solution
($\sinh\alpha=1$, $\beta=\infty$). Plotting both these classes of $P=Q$ extreme
solutions on one graph gives a piecewise continuous curve, see Fig.~3., which
is just
the slice $P=Q$ through the surface in Fig.~2.

\begin{figure}
\epsffile{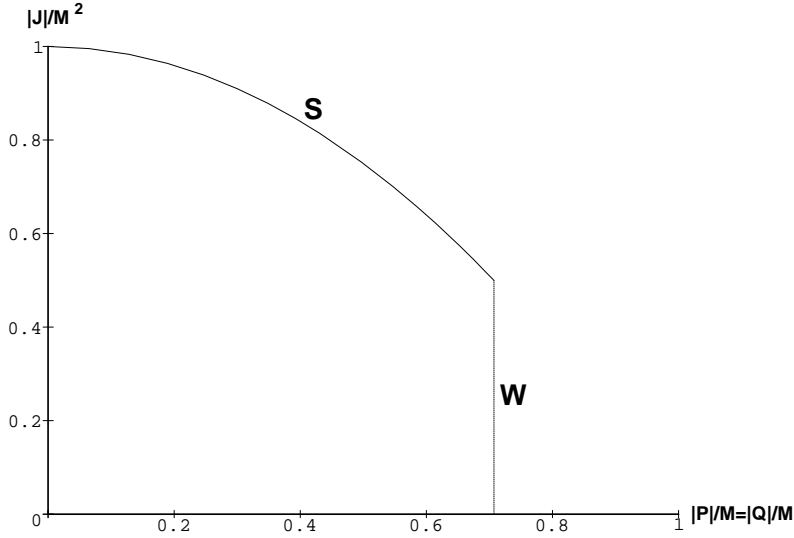}
\caption{Extreme $P=Q$ Solutions}
\end{figure}

\sect{Classical Properties}

It will be useful here to calculate some of the classical quantities associated
with
the general rotating solution (\ref{GenMetric}). First, unlike the non-rotating
solutions, (\ref{GenMetric}) has electric and magnetic dipole moments which are
respectively
\begin{equation}
D = {PJ\left(M+\Sigma/\sqrt{3}\right)\over\left(M-\Sigma/\sqrt{3}\right)^2-P^2}
\end{equation}
and
\begin{equation}
\mu =
{QJ\left(M-\Sigma/\sqrt{3}\right)\over\left(M+\Sigma/\sqrt{3}\right)^2-Q^2}.
\end{equation}
Therefore the gyromagnetic and gyroelectric ratios are
\begin{equation}
g_{_M} =
{2M\left(M-\Sigma/\sqrt{3}\right)\over\left(M+\Sigma/\sqrt{3}\right)^2-Q^2}
\label{GyroM}
\end{equation}
and
\begin{equation}
g_{_E} =
{2M\left(M+\Sigma/\sqrt{3}\right)\over\left(M-\Sigma/\sqrt{3}\right)^2-P^2}.
\label{GyroE}
\end{equation}
Note that the discrete electromagnetic duality transformation (\ref{e-m}) is
equivalent to exchanging $P$ and $Q$ in (\ref{GenMetric}) and also changing the
signs
of $J$ and $\Sigma$. This also exchanges $g_{_M}$ and $g_{_E}$. The purely
electric
solutions can be obtained by the Lorentz boost (\ref{LBoost}) applied to the
Kerr
solution which is equivalent to setting $\beta=0$ in (\ref{NewChi}) and
(\ref{Boost}).
This gives a gyromagnetic ratio of $2-v^2$ in agreement with \cite{GibWil}. The
gyroelectric ratio of the dual $Q=0$ solutions is then also $2-v^2$ where now
$\gamma=(1-v^2)^{-{1\over 2}}=\cosh\beta$. When $P=Q$, $\Sigma=0$ and the
gyromagnetic
and gyroelectric ratios are equal:
\begin{equation}
g_{_M} = g_{_E} = {2M^2\over M^2-Q^2} = {2M^2\over M^2-P^2}.
\end{equation}
Thus these 2 ratios range from 2 up to 4 in the extreme case. In general it
turns out
that $g_{_M}$ and $g_{_E}$ are bounded below by 1, with $g_{_M}=1$ only for the
extreme electric (plane wave) solution and $g_{_E}=1$ only for the \GPS\
magnetic
monopole.

In the more general dyonic case, however, (\ref{GyroM}) and (\ref{GyroE}) show
that
$g_{_M}$ and $g_{_E}$ are not bounded above at all but can become arbitrarily
large.
For example, as $P\rightarrow 0$ and $Q\rightarrow 2M$,
$\Sigma\rightarrow\sqrt{3}M$
and $g_{_E}\rightarrow\infty$. This is to be contrasted with the Kerr-Newman
dyonic
black holes of Einstein-Maxwell theory which all have $g_{_M}=g_{_E}=2$. The
unusual
behaviour of the gyromagnetic and gyroelectric ratios in \KK\ theory only
becomes
apparent when rotating dyons are considered. This may also be the case for
other
values of the dilaton coupling constant $b$.

(\ref{GenMetric}) is stationary and axisymmetric, therefore
$\xi={\partial\over\partial t}$ and $m={\partial\over\partial\phi}$ are Killing
vector
fields. The outer event horizon, given by the larger root of $\Delta$,
$r=r_{+}$ where
\begin{equation}
r_\pm = M \pm\sqrt{M^2+\Sigma^2-P^2-Q^2-a^2}
\end{equation}
is a Killing horizon of the Killing vector field
\begin{equation}
k = {\partial\over\partial t} + \Omega_H{\partial\over\partial\phi}
\end{equation}
where $\Omega_H$ has the interpretation of the angular velocity of the event
horizon.
The requirement that $k$ be null on the horizon $H$ gives
\begin{equation}
\Omega_H = -\left.{1\over{\omega^0}_\phi}\right|_H = {a^2\over 2J}\left[r_{+} -
M +
{\left(M^2+\Sigma^2-P^2-Q^2\right)\left(M+\Sigma/\sqrt{3}\right)\over
\left(M+\Sigma/\sqrt{3}\right)^2-Q^2}\right]^{-1}.
\label{Omega}
\end{equation}

In general the solution will possess an ergoregion which will be given by the
region
between $r=r_{+}$ and the larger zero of $f^2$:
\begin{equation}
r = M  + \sqrt{M^2+\Sigma^2-P^2-Q^2-a^2\cos^2\theta}.
\end{equation}
The extreme solutions lying on the wall {\bf W} in Fig.~2., however, have $a=0$
and
therefore they will have no ergoregion and zero angular velocity even though
they may
have non-zero ADM angular momentum. In Einstein-Maxwell theory it is believed
that a
non-rotating black hole solution (in the sense that $k=\xi$, i.e.\
$\Omega_H=0$) must
be static and spherically symmetric \cite{Carter}. The extreme dyons of \KK\
theory on
{\bf W}, however, may be non-rotating in this sense and still be neither static
nor
spherically symmetric. (\ref{PeqQJneq0}) is the simplest example of such a
solution.
Using (\ref{Ja}) it can be seen that only those solutions with $(P/M)^{2\over
3}+(Q/M)^{2\over 3}=2^{2\over 3}$ can have this unusual property, since only
for these
solutions is it possible for $a$ to be zero when $J$ is non-zero.

The surface gravity of the event horizon  is
\begin{equation}
\kappa = {r_{+}-M\over\left.\sqrt{AB}\right|_{r=r_{+},\theta=0}}
\label{Kappa}
\end{equation}
and so this vanishes in the extreme limit (on {\em both} surfaces {\bf S} {\em
and}
{\bf W}).

The area of the event horizon given by
\begin{equation}
{\cal A} = \int_{\theta=0}^\pi d\theta \int_{\phi=0}^{2\pi}d\phi
\left.\sqrt{g_{\theta\theta}g_{\phi\phi}}\right|_{r=r_{+}}
\end{equation}
is
\begin{equation}
{\cal A} = {8\pi J\over a}\left[r_{+} - M +
{\left(M^2+\Sigma^2-P^2-Q^2\right)\left(M+\Sigma/\sqrt{3}\right)\over
\left(M+\Sigma/\sqrt{3}\right)^2-Q^2}\right].
\end{equation}

We define the co-rotating electrostatic potential by
\begin{equation}
\Phi = k.A = A_t + \Omega_H A_\phi.
\end{equation}
This is a gauge dependant quantity. In the gauge chosen in (\ref{Aem}) and
(\ref{C})
$\Phi\rightarrow 0$ as $r\rightarrow\infty$ and on the horizon $\Phi=\Phi_H$
where
\begin{equation}
\Phi_H = \left.{-{\omega^5}_\phi\over 2{\omega^0}_\phi}\right|_H.
\label{PhiH}
\end{equation}
In the electrically charged, magnetically neutral case this leads to a
Smarr-type
formula \cite{Hawk}
\begin{equation}
M = {\kappa{\cal A}\over 4\pi} + 2\Omega_HJ + \Phi_HQ.
\label{Smarr1}
\end{equation}
In the magnetically charged, electrically neutral case one can use
electromagnetic
duality to define a co-rotating magnetostatic potential $\Psi$ which may be
obtained
from $\Phi$ by exchanging $P$ and $Q$ and changing the signs of $J$ and
$\Sigma$. The
Smarr formula will then simply be
\begin{equation}
M = {\kappa{\cal A}\over 4\pi} + 2\Omega_HJ + \Psi_HP.
\label{Smarr2}
\end{equation}
In the dyonic case it turns out that the mass obeys the obvious generalization
of
(\ref{Smarr1}) and (\ref{Smarr2})
\begin{equation}
M = {\kappa{\cal A}\over 4\pi} + 2\Omega_HJ + \Phi_HQ + \Psi_HP.
\label{Smarr}
\end{equation}
Therefore $\Phi_HQ$ and $\Psi_HP$ may be interpreted as the contributions to
the total
energy from the electric and magnetic charges respectively. Now $M$ is a
homogeneous
function of degree 1 in the variables $A^{1\over 2}$, $J^{1\over 2}$, $Q$ and
$P$ and
so using Euler's theorem (\ref{Smarr}) leads to the generalized first law of
black
hole thermodynamics
\begin{equation}
dM = {\kappa\over 8\pi}d{\cal A} + \Omega_HdJ + \Phi_HdQ + \Psi_HdP.
\end{equation}

\sect{Stability}

Extreme black holes often behave like elementary particles and so it is
interesting to
investigate their stability, i.e.\  whether it is possible for them to split
into
smaller black holes. In Einstein-Maxwell theory this is forbidden by energy
conservation. However, it has been pointed out \cite{KalLin} that in the case
of
extreme dilatonic black holes with $b=1$ the relation between $M$, $P$ and $Q$
is such
that energy conservation no longer forbids splitting of the black holes. It
might also
be argued that the second law of black hole thermodynamics would prevent such
splitting. However, Kallosh et al.\ \cite{KalLin} showed that the entropy $S$
vanishes
for extreme dilaton black holes with $b=1$ and later work \cite{HawHor},
\cite{Teit}
has shown that $S=0$ for {\em all} extreme black holes, so the second law does
not
forbid the splitting of extreme black holes \cite{GibKal}. Finally, in
classical
General Relativity the area law prevents extreme black holes from splitting but
in
a full quantum  theory of gravity the emission of sufficiently small extreme
black
holes may be allowed. In this section we will only investigate the question of
whether
it is energetically favourable for extreme black holes to split.

Consider first the non-rotating extreme solutions of some general theory with
conserved charges $Q_1,Q_2,\dots Q_n$ each having units of mass. The mass will
be a
homogeneous function of degree 1 of the charges. Hence

\medskip
\,\,\,\,(i).\,\,\,\,$M(\mbox{\boldmath $x$})\ge 0 \quad {\rm and} \quad
M(\mbox{\boldmath $x$})=0 \Leftrightarrow \mbox{\boldmath $x$}=0$

\medskip
\,\,\,\,(ii).\,\,\,$M(\lambda\mbox{\boldmath $x$}) = \lambda M(\mbox{\boldmath
$x$})
\quad \forall\lambda\in I\!\!R$

\bigskip
\noindent
where $\mbox{\boldmath $x$}=(Q_1,Q_2,\dots Q_n)\in I\!\!R^n$. A solution
specified by
$\mbox{\boldmath $x$}\in I\!\!R^n$ may be considered to be unstable if it is
energetically favourable for it to decay into 2 new solutions labelled by
$\mbox{\boldmath $x_1$},\mbox{\boldmath $x_2$}\in I\!\!R^n$ with
$\mbox{\boldmath
$x_1$}+\mbox{\boldmath $x_2$}=\mbox{\boldmath $x$}$. The condition for all
solutions
to be stable is therefore

\medskip
\,\,\,\,(iii).\,\,$M(\mbox{\boldmath $x_1$}+\mbox{\boldmath $x_2$}) \le
M(\mbox{\boldmath $x_1$}) + M(\mbox{\boldmath $x_2$}) \quad
\forall\mbox{\boldmath
$x_1$},\mbox{\boldmath $x_2$}\in I\!\!R^n.$

\bigskip
\noindent
The 3 conditions (i), (ii) and (iii) may be recognized as the conditions for
$M$ to
define a norm on $I\!\!R^n$, the stability condition being the triangle
inequality. A
convenient way of identifying stable mass functions is then given by the fact
that the
unit ball of a norm on $I\!\!R^n$ is a convex subset of $I\!\!R^n$. To see this
consider $\mbox{\boldmath $x_1$},\mbox{\boldmath $x_2$}$ in the unit ball of a
norm
$M$ on $I\!\!R^n$, $M(\mbox{\boldmath $x_1$}),M(\mbox{\boldmath $x_2$})\le 1$.
Then
$M(\lambda\mbox{\boldmath $x_1$}+(1-\lambda)\mbox{\boldmath $x_2$})\le\lambda
M(\mbox{\boldmath $x_1$})+(1-\lambda)M(\mbox{\boldmath $x_2$})\le 1$. So
$\lambda\mbox{\boldmath $x_1$}+(1-\lambda)\mbox{\boldmath $x_2$}$ is also in
the unit
ball. Hence the unit ball is convex.

In the case of Einstein-Maxwell theory coupled to a dilaton field, the unit
balls of
the mass functions are given by the curves in Fig.~1. Hence the extreme
non-rotating
dyons are stable for dilaton couplings $b\le 1$ and unstable for $b>1$. Another
way of
seeing this is to use Minkowski's inequality which says that
$$M(Q,P) = {1\over\sqrt{1+b^2}} \left(|Q|^n+|P|^n\right)^{1\over n}$$
satisfies the triangle inequality for $n\ge 1$, i.e.\  for $b\le 1$. Since \KK\
theory
has $b=\sqrt{3}>1$, its extreme non-rotating dyons are unstable.

The rotating dyons with $J\le PQ$ on the surface {\bf W} of Fig.~2. have the
same
relation between $M$, $Q$ and $P$, independent of $J$, so they will be
unstable. In
the more general case, however, $M$ will be a function of $J$ as well and the
requirement for stability becomes
\begin{equation}
M(Q_1+Q_2,P_1+P_2,J_1+J_2) \le M(Q_1,P_1,J_1) + M(Q_2,P_2,J_2).
\end{equation}
Now $J$ has units of (mass)$^2$ and so $M$ is no longer a homogeneous function
and the
argument used in the non-rotating case will not work. It is possible, however,
to
write $J$ uniquely as a function of $M$, $Q$ and $P$ (at least up to a sign)
and $J$
will then be a homogeneous function of degree 2. Writing $\mbox{\boldmath
$x$}=(M,Q,P)$, the stability condition becomes
\begin{equation}
J(\mbox{\boldmath $x_1$}+\mbox{\boldmath $x_2$}) \ge J(\mbox{\boldmath $x_1$})
+
J(\mbox{\boldmath $x_1$}).
\end{equation}
Note that the inequality has become reversed. Hence if the solutions are {\em
unstable} the unit ball $J(\mbox{\boldmath $x$})\le 1$ will be a convex region
of
($M$,$Q$,$P$) space. For extreme rotating solutions on the surface {\bf S} of
Fig.~2.,
the boundary of the unit ball in ($M$,$Q$,$P$) space is given by setting $J=1$
which
gives
$$M = {\cosh\alpha\left(1+\cosh^2\alpha\cosh^2\beta\right)\over
2\sqrt{\cosh\beta}\left(1+\sinh^2\alpha\cosh^2\beta\right)^{3\over 4}}$$
\begin{equation}
Q = {\sinh\alpha\left(1+\sinh^2\alpha\cosh^2\beta\right)^{1\over 4}\over
\sqrt{\cosh\beta}}
\end{equation}
$$P =
{\sinh\beta\sqrt{\cosh\beta}\over\left(1+\sinh^2\alpha\cosh^2\beta\right)^{3\over
4}}.$$
A plot of this surface shows it to be concave and this is confirmed by the
determinant
of the matrix of second derivatives
\begin{equation}
\left|
\begin{array}{cc}
{\partial^2M\over\partial Q^2} & {\partial^2M\over\partial Q\partial P} \\
 & \\
{\partial^2M\over\partial P\partial Q} & {\partial^2M\over\partial P^2}
\end{array}
\right|
= {\left(1+\sinh^2\alpha\cosh^2\beta\right)^{3\over 2}\over
2\cosh^4\alpha\cosh\beta\left(3\cosh^2\alpha\cosh^2\beta-1\right)}
\end{equation}
which is everywhere positive. So the unit ball of $J(M,Q,P)$ is concave, hence
the
rotating dyons on the surface {\bf S} of Fig.~2. are stable.

\sect{Thermodynamic Quantities}

The thermodynamic quantities of (\ref{GenMetric}) calculated earlier are more
easily
expressed in terms of the 4 independent variables $M_K,a,\alpha,\beta$. The
temperature defined by $T={\kappa\over 2\pi}$ and using (\ref{Kappa}) is
\begin{equation}
T = {1\over 4\pi}{\sqrt{M_K^2-a^2} \over M_K\cosh\beta \left[M_K\cosh\alpha +
\sqrt{M_K^2-a^2}\sqrt{1+\sinh^2\alpha\cosh^2\beta}\right]}
\end{equation}
which, as expected, vanishes for extreme solutions on {\bf S} and {\bf W} since
$|a|=M_K$ for these solutions. The entropy $S={1\over 4}{\cal A}$ is
\begin{equation}
S = 2\pi M_K\cosh\beta \left[M_K\cosh\alpha +
\sqrt{M_K^2-a^2}\sqrt{1+\sinh^2\alpha\cosh^2\beta}\right].
\label{Entropy}
\end{equation}
The angular velocity of the event horizon (\ref{Omega}) is
\begin{equation}
\Omega_H = {a \over 2M_K\cosh\beta \left[M_K\cosh\alpha +
\sqrt{M_K^2-a^2}\sqrt{1+\sinh^2\alpha\cosh^2\beta}\right]}.
\end{equation}
The angular momentum is
\begin{equation}
J = aM_K\cosh\beta\sqrt{1+\sinh^2\alpha\cosh^2\beta}.
\end{equation}
The electrostatic potential on the horizon (\ref{PhiH}) is
\begin{equation}
\Phi_H = { \sinh\alpha \left[M_K +
\sqrt{M_K^2-a^2}{\cosh\alpha\cosh^2\beta\over
\sqrt{1+\sinh^2\alpha\cosh^2\beta}}\right] \over 2\left[M_K\cosh\alpha +
\sqrt{M_K^2-a^2}\sqrt{1+\sinh^2\alpha\cosh^2\beta}\right]}.
\end{equation}
The electric charge is
\begin{equation}
Q = M_K\sinh\alpha \sqrt{1+\sinh^2\alpha\cosh^2\beta}.
\end{equation}
The magnetostatic potential on the horizon is
\begin{equation}
\Psi_H = { \sinh\beta \left[M_K +
\cosh\alpha\sqrt{M_K^2-a^2}\sqrt{1+\sinh^2\alpha\cosh^2\beta}\right] \over
2\cosh\beta
\left[M_K\cosh\alpha +
\sqrt{M_K^2-a^2}\sqrt{1+\sinh^2\alpha\cosh^2\beta}\right]},
\end{equation}
and the magnetic charge is
\begin{equation}
P = {M_K\sinh\beta\cosh\beta \over \sqrt{1+\sinh^2\alpha\cosh^2\beta}}.
\end{equation}
Then the Smarr formula becomes
\begin{equation}
M = 2TS + 2\Omega_HJ + \Phi_HQ + \Psi_HP
\end{equation}
and the generalized first law is
\begin{equation}
dM = TdS + \Omega_HdJ + \Phi_HdQ + \Psi_HdP.
\end{equation}
In the previous section it was shown that the only extreme solutions which
could be
unstable on energetic grounds are those on the surface {\bf W}. It was argued
in
\cite{HawHor} and \cite{Teit} that the entropy of extreme black holes is zero
and so
(\ref{Entropy}) does not apply to extreme solutions. Thus the second law does
not
forbid the splitting of extreme black holes on {\bf W}. It turns out that
(\ref{Entropy}) actually does give $S=0$ for extreme solutions on {\bf W} (but
not on
{\bf S}) and so the event horizons of solutions on {\bf W} have zero surface
area.
Thus the area law does not forbid the splitting of extreme black holes on {\bf
W}
either and these solutions are expected to be genuinely unstable.

\sect{Conclusions}

In this paper we have given an explicit form for the most general dyonic black
hole
solution of the 5 dimensional \KK\ theory. The formulae in terms of the
physical
parameters $M$,$P$,$Q$,$J$ are fairly complicated and for many purposes, such
as
thermodynamics, it is more convenient to use the parameters
$M_K$,$a$,$\alpha$,$\beta$. The solution (\ref{GenMetric}) reduces to the known
non-rotating solutions \cite{Leut}--\cite{GibWil} on setting $a=J=0$. The
magnetically
neutral case $\beta=P=0$ is simply the Lorentz boosted Kerr solution
\cite{GibWil},
\cite{HorHor}, \cite{FroZel}. The electrically neutral, magnetically charged
solutions
$\alpha=Q=0$ are the electromagnetic duals of the magnetically neutral,
electrically
charged solutions.

The gyromagnetic and gyroelectric ratios of the black holes in this theory have
been
calculated (\ref{GyroM}), (\ref{GyroE}) and they were found to be bounded below
by 1
but not bounded above. This is a surprising result since the black holes of
Einstein-Maxwell theory (the Kerr-Newman solutions) and those of heterotic
string
theory found by Sen \cite{Sen} all have gyromagnetic and gyroelectric ratios
bounded
above by 2. (\cite{Sen} only gives the gyromagnetic ratio for the electrically
charged
case. It would be interesting to see what happens to this ratio in the more
general
dyonic case.)

The surface of extreme solutions was found to be made up of 2 distinct surfaces
{\bf
S} and {\bf W} with $J>PQ$ and $J<PQ$ respectively. The 2 surfaces meet at a
line
given by \hbox{$(P/M)^{2\over 3}+(Q/M)^{2\over 3}=2^{2\over 3}$}, $J=PQ$. Thus
the
full surface of extreme solutions is no longer a smooth surface, as it is in
Einstein-Maxwell theory. The solutions on the 2 surfaces behave completely
differently. Those on {\bf S} (which includes the extreme Kerr solution) show
all the
normal characteristics of rotating solutions, such as non-zero angular velocity
of the
event horizon and an ergoregion outside the event horizon. Those extreme
solutions on
{\bf W}, however, behave more like non-rotating solutions since their event
horizons
have zero angular velocity and they have no ergoregion whilst they are not
spherically
symmetric and they have non-zero ADM angular momentum.

The stability of the extreme solutions viewed as elementary particles was
investigated. It was found that extreme dyons on the surface {\bf S} of Fig.~2.
with
$J\ge PQ$ are stable, whereas those on the surface {\bf W} with $J<PQ$ are
unstable.
The solutions on {\bf S} with $J>PQ$ have $a\ne 0$, non-zero angular velocity
and they
have ergoregions and so they are expected to lose angular momentum by
super-radiance.
As the angular momentum decreases, however, and one moves down the surface {\bf
S}
towards {\bf W}, $J\rightarrow PQ$, $a\rightarrow 0$, the ergoregion vanishes
and the
super-radiance is turned off. Therefore the extreme solutions never quite lose
enough
angular momentum to reach {\bf W} and become unstable, instead they
asymptotically
approach the $J=PQ$ curve.

\bigskip
\hfil{\large\bf Acknowledgements}\hfil

\medskip
\noindent
The author would like to thank G. Gibbons for many stimulating discussions and
helpful
comments in preparing this paper. This work was supported by EPSRC grant no.\
9400616X.

\end{document}